# Techno-Economic Assessment Models for 5G

**Carlos Bendicho** [1]

Independent ICT Researcher

ACM Member, IEEE Communications Society Member, FITCE Member

https://orcid.org/0000-0002-8538-0043

**Abstract**

This paper proposes the characteristics a techno-economic model for 5G should have considering both mobile network operator´s perspective and end user´s needs. It also presents a review and classification of models in the literature based on the characteristics of such theoretical techno-economic reference model. The performed analysis identifies current gaps in the techno-economic modeling literature for 5G architectures and shows it can be enhanced using agile techno-economic models like the Universal Techno-Economic Model (UTEM) created and developed by the author to industrialize assessment of different technological solutions, considering all market players perspectives and applicable to decision-making in multiple domains. This model can be used for an effective and agile 5G techno-economic assessment, including not only network deployment perspective but also customers´ and end users´ requirements as well as other stakeholders´ to select the most adequate 5G architectural solution considering both technical and economic feasibility. UTEM model is currently available for all industry stakeholders under specific license of use.

## I. Introduction

The deployment of any new access technology such as 5G, requires careful planning and, above all, the selection of the optimal technological solution from a techno-economic perspective, i.e.: considering both technical feasibility and economic viability not only from a Total Cost of Ownership (TCO) perspective, but also for business viability regarding Net Present Value (NPV) and Internal Rate of Return economic parameters. This document contemplates the redefinition of techno-economic models as those considering both technical and economic feasibility of complex technical systems [1, 2], instead of the traditional definition that only takes into account economic feasibility [3].

---

[1] Dr. Carlos Bendicho holds M.Sc. and Ph.D. degrees in Telecommunications Engineering from Bilbao School of Engineering, University of the Basque Country, Spain. He is also MBA from Technical University of Madrid and MIT Sloan Executive Program in Artificial Intelligence and Strategy, Massachusetts Institute of Technology.

5G system is assumed to be integrated by multiple Radio Access Technologies (RATS), such as evolved LTE, Wi-Fi, New Radio (NR), WiMAX, etc., with heterogeneous deployments rather than homogeneous base station placements [4]. 5G also uses low-band, mid-band and higher carrier frequencies (mm-Wave), whose diverse propagation conditions will require those heterogeneous deployments for adequate coverage. 5G uses Massive MIMO (Multiple Input Multiple Output) technology to increase number of antennas in base stations and terminal devices, leveraging Space Division Multiple Access (SDMA) so that higher bandwidths are possible. Main service types defined in IMT-2020 by ITU-R are enhanced Mobile Broad Band (eMBB), massive Machine-Type Communications (mMTC) and Ultra Reliable Low Latency Communications (URLLC) [4].

Therefore, 5G system increases complexity vs previous access technologies, which makes techno-economic evaluation of 5G solutions a challenging task. Hence, there is a need for solvent and agile 5G techno-economic assessment models.

However, techno-economic models in the literature are elaborated for specific scenarios and lack generalization capabilities that allow their adaptation to different use cases and evolving 5G architectures. Besides, they are not flexible enough to integrate new technical and economic parameters as well as all industry stakeholders´ perspectives, and do not allow agile assessment through automation. Therefore, it makes sense to research for new 5G techno-economic assessment models that cover these needs.

This paper shows in Section II a review of relevant models in the literature and the research problem in Section II.C. Section III proposes the characteristics of a theoretical 5G techno-economic assessment reference model considering both mobile network operators´ perspective and end users´ needs. Section IV presents a classification of models in the literature according to the degree of compliance with the characteristics of the theoretical model, identifying current gaps. Eventually, Section V shows conclusions and future work about using a universal techno-economic model created and developed by the author to reduce identified gaps and effectively assess 5G solutions.

## II. Review of Techno-Economic Models in the literature

This section presents a review of relevant 5G techno-economic assessment papers. The selection of papers has been made by searching "5G Techno economic" in IEEE Xplore, Google Scholar, Scopus and Springer, Elsevier databases. The following section II.A shows a review of those relevant articles and section II.B presents conclusions of this literature review.

### A. Models in the literature

This section presents the review of 18 relevant papers about 5G techno-economics in the literature.

Reference [5] shows a techno-economic framework for 5G transport networks distinguishing homogeneous deployment of macro-cells and heterogeneous deployment (HetNet) mixing macro-cells with indoor coverage and selecting three alternatives for back hauling: (1) microwave, (2) leased fiber (3) owned fiber. It provides TCO and NPV economic outputs and concludes that best back hauling strategy is leased fiber with both heterogeneous and homogeneous deployments.

Reference [4] presents the formal technical evaluation framework for 5G: IMT-2020 framework by ITU-R to evaluate 5G radio interface technologies (RIT) or set of RITs (SRIT), comprising the three main service types: eMBB, mMTC and URLLC. It does not consider any economic parameters for input or output.

Several 5G PPP (5G Private Partnership Projects) projects have investigated the topic of 5G technical parameters assessment, constituting the 5G PPP Evaluation framework that considers the following KPIs (Key Performance Indicators) [4]:

- 6 inspection KPIs:
  - Bandwidth and channel bandwidth scalability (at least supporting 1 GHz)
  - Deployment in IMT bands, at least on one.
  - Operation above 6 GHz
  - Spectrum flexibility: capability to accommodate different downlink (DL) and uplink (UL) transmission patterns in both paired and unpaired frequency bands.
  - Inter-system handover between 5G and at least one legacy system
  - Efficient support of a wide range of services over a continuous single block of spectrum.
- 5 Analytical KPIs:
  - Mobility interruption time: capability to provide a continuous connectivity for devices on the move. 0 ms interruption time is possible, if multi connectivity solutions are used as required by ITU-R.
  - Peak data rate: the highest theoretical single-user data rate. ITU-R requires >=20 Gbps for DL and >=10 Gbps for UL.
  - mMTC device energy consumption: single 5Wh battery. Ability to provide an energy efficient procedure for IoT reaching >= 10 years lifetime.
  - Control plane latency: <= 20ms
  - User plane latency: <= 1ms
- 10 Simulation KPIs:
  - Experienced user throughput
  - Traffic volume density
  - E2E latency
  - Reliability
  - Retainability
  - mMTC device density
  - RAN energy efficiency
  - Supported velocity
  - Complexity
  - Coverage

Simulation KPIs performance depends on the use case or scenario under study, e.g.: dense urban information society, connected cars, massive deployment of sensors and actuators, virtual reality office or broadband access everywhere are use cases considered in METIS-II project [6].

Reference [4] contributes also with an economic assessment with OPEX, CAPEX, Revenues, ARPU and NPV as output parameters.

Reference [7] proposes a techno-economic assessment framework for 5G deployment with a capacity assessment module and a demand assessment module that provides capacity, coverage, TCO, NPV and energy efficiency outputs considering multi operator sharing strategies in both RAN and back hauling with fibre.

Reference [8] shows techno-econonomic evaluation of mm-Wave RANs, considering both Distributed RAN (DRAN) as depicted in Fig. 1, and different degree of RAN centralization or Cloud RAN (CRAN) approaches with owned or leased fiber transport network (front haul and back haul) as shown in Fig. 2. This reference also addresses capacity and coverage analysis of "immersive 5G experience" with advanced service types like Immersive Reality or Ultra High Definition (UHD) Video within mmMAGIC 5G PPP project.

Although outside the scope of this work, [9] deserves special mention as it introduces future scenarios planning of Industrial 5G, using Value Network Configuration (VNC) in order to identify different architectural components, business actors and their roles, which can be complementary to ulterior techno-economic assessment.

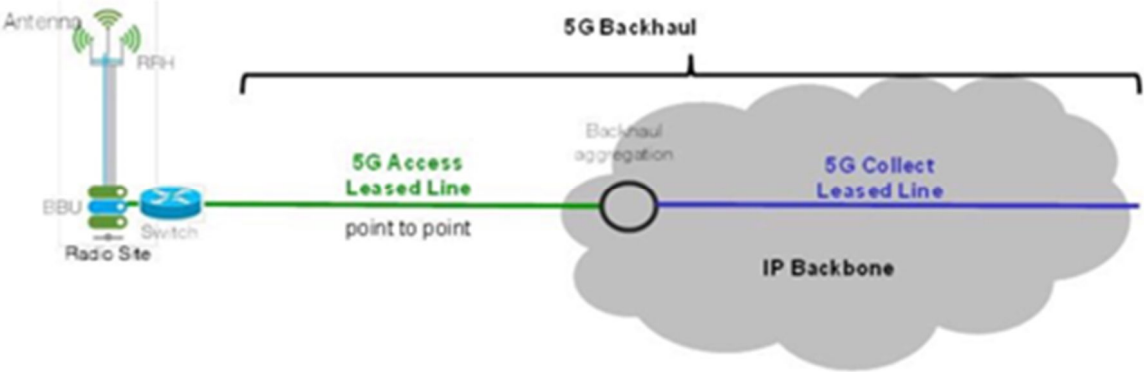

*Fig. 1. Distributed RAN (DRAN) with leased lines [8].*

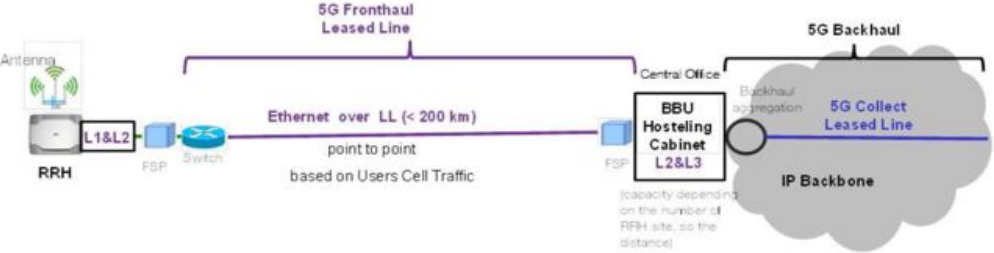

*Fig. 2. Centralized RAN (CRAN) in split E with leased lines [8].*

In [10], authors show a cost-benefit model to perform an analysis on how to deploy 5G technology over existing 4G networks. It analyses coverage, capacity, price and cost (CAPEX, OPEX) for different scenarios using different type of base stations (Macro base stations, Micro Base Stations, Pico Base Stations, Femto base stations) and IEEE 802.11ac Access Points. They also determine optimal price having into account elasticity of demand (PED) with a price of $1.75 per GB/user/month for Shanghai (China) and 100GB/user/month of demanded data volume reference, providing in that case 57,1% of profit margin. This cost-benefit model considers that if ROI of a 5G deployment scenario is higher than its TCO, that deployment is viable.

Reference [11] outlines a conceptual cost-capacity model to analyse 5G mm-Wave systems, considering heterogeneous deployments, and taking into account estimated profit margins to make decisions. The model is applied to "virtual reality office" use case and assesses deployment alternatives with 4G-LTE-A Macro Base Stations placed outside the buildings of a new office center of 1.0 km2 and 5G mmW Pico Base Stations sites and WLAN IEEE 802.11ad inside buildings.

Regarding different verticals within 5G PPP ONE5G project, [12] provides techno-economic evaluations on 5GNR network deployments to cover 4 verticals: Automotive (V2X), Smart Cities (with NB-IoT and LTE-M), Long range connectivity in rural areas, and Disaster and Emergency support for airborne emergency crews using fleets of drones for 5G connectivity. This reference evaluates distributed RAN and centralized RAN (CRAN) approaches considering fronthaul and backhaul deployments based on leased fiber, owned fiber or owned microwave. It includes, for Automotive vertical, the use of Multi Access Edge Computing nodes (MEC) to fulfill low latency, high availability and high reliability requirements.

Reference [13] presents a techno-economic analysis of several alternatives: ultra-dense deployments using small cells (picocells and femtocells), macro cells and Distributed Antenna Systems (DAS) alone or in combination with macro cells, providing CAPEX, OPEX, IMPEX (implementation expenses) and TCO.

In [14], authors introduce a techno-economic model to estimate CAPEX, OPEX and TCO of 5G network architecture based on Centralized RAN or Cloud RAN composed by virtualized Base Stations or Software Defined Base Stations on Edge Data Centers commodity servers and virtualized Evolved Packet Core (vEPC). This architecture is oriented to leverage the potential cost reduction offered by the combination of SDN/NFV technologies.

Reference [15] presents a methodology framework for techno-economic evaluation of 5G-PICTURE 5G-PPP project architectures considering different transport network technologies: mmWave for Backhauling (BH) and Fronthauling (FH) links, Sub-6 GHz transceivers for Backhauling, Point-to-Point optical connections for BH/FH links, and other technologies like WDM-PON (Wavelength Division Multiplexing – Passive Optical Networks). Access transport links can be either multiplexed by ITU-T G.998.4 optical connections at first level at street cabinets or aggregated directly at second level at Central Offices. Third level transport uses TSON (Time Shared Optical Network) as depicted in Fig. 3. It also considers the mix of different functional splits in the network deployment (eCPRI Split A, eCPRI B, eCPRI C, eCPRI D, eCPRI ID, eCPRI IID/IU, eCPRI E). This evaluation framework provides TCO, CAPEX and OPEX of different deployment alternatives between the access network nodes (RRH, Macro-sites, Small Cells) and the core network segment, including compute, storage and software resources of MEC (Mobile Edge Computing), vBBU (virtual Base Band Units) and core network components. It uses a macroscopic techno-economic analysis to select the most cost-efficient and applicable scenario, which is then analyzed using a microscopic analysis regarding detailed information about locations, sites and vertical applications.

In [16], a Cognitive Radio architecture is suggested based on femtocells, in order to share spectrum between a primary network with primary users that buy and pay for licensed spectrum and secondary networks with secondary femtocell users aware of the radio channel conditions. For each femtocell, there is a Base Station that provides services to secondary users whenever primary users are not using allocated bandwidth resources. It uses Stackelberg game as game theoretic approach in economics to help reducing costs and finding equilibrium among parameters and participants in order for them to obtain maximum profit. It provides CAPEX, OPEX, TCO, Revenues and Profit comparison between proposed Cognitive Radio architecture and SDN architectural approach introduced in [14].

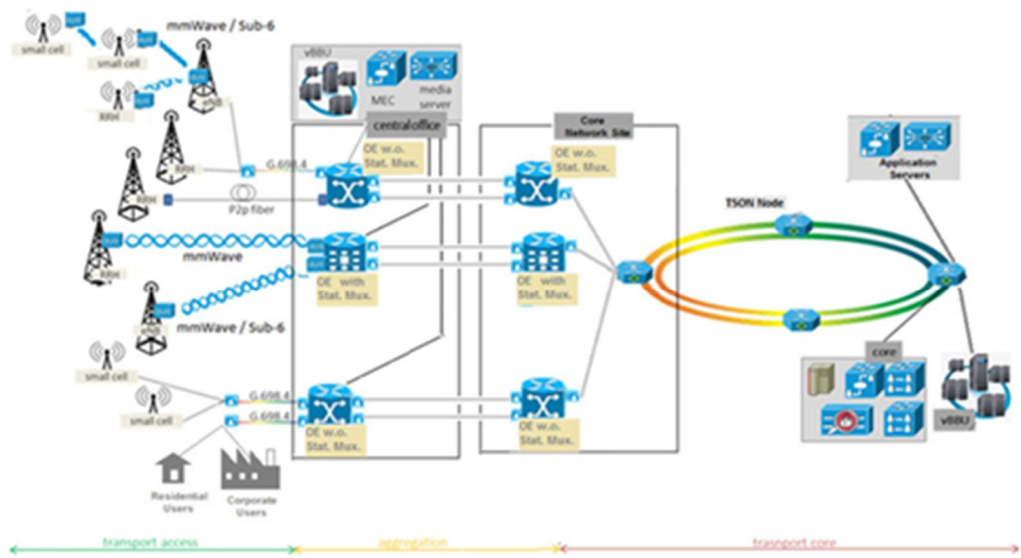

*Fig. 3. Network deployment model and technologies in 5G-PICTURE project [15].*

Reference [17] shows economic comparison of four technologies for mm-Wave: Distributed Antenna System (DAS); Modified DAS including NFV virtualization of outdoor and indoor antennas, bandwidth and base station, thus reducing OPEX; Multiple Input Multiple Output (MIMO) with several antennas on the transceiver`s and receiver`s sides (2, 4, 8, 16); and Massive MIMO which outnumbers MIMO considering 64, 128, 256, 512 antennas. It provides CAPEX, OPEX and TCO economic outputs.

In [18] an integrated satellite-cellular system for 5G and beyond (B5G) is presented under a KR-EU (Korean-European Union) joint project called 5G-ALLSTAR, although no techno-economic model is shown and therefore it is outside of the scope of this paper, but the reader can get deep knowledge of this hybrid system to develop Multi-Connectivity (MC) technology for seamless, reliable and ubiquitous broadband services with 5G-NR cellular system and NR-based satellite access.

Reference [19] introduces a Technical Rate (TR) to represent the technical improvement when transitioning between different scenarios on a roadmap for a full 5G deployment for eMBB, Massive IoT (MIoT) and Mission Critical Service (MCS) use cases. It considers three scenarios shown in Fig. 4: Scenario A: LTE-A Pro system composed by enhanced RAN and EPC (Evolved Packet Core); Scenario B: Both New Radio (NR) partially deployed (possible due to slicing) and LTE RAN will be connected to the two cores: Next Generation Core Network (NGCN) and EPC that will interwork; Scenario C: NR RAN interworking with legacy RATs (Radio Access Technologies) and NGCN interworking with EPC (IMT-2020 compliant).

Technical rate in [19] is a weighted measure of KPIs improvement and is aligned with techno-economics redefinition made by the present paper´s author in [1, 2] that emphasizes not only economic viability evaluation but also technical feasibility evaluation of complex technical systems. Reference [19] includes also a Recommended Score to show which use cases could work in each scenario, as well as a Sensitivity Analysis (SA).

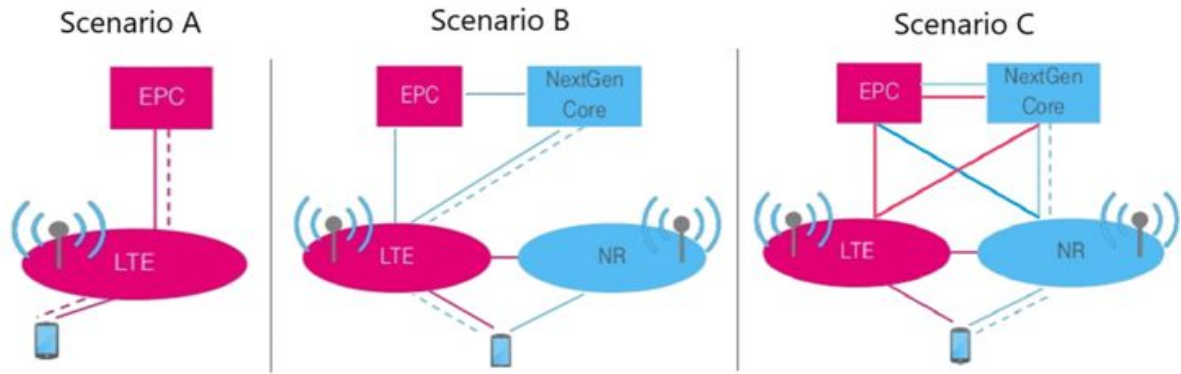

*Fig. 4. Deployment scenarios considered in [19]: Scenario A: based on LTE-A Pro, Scenario B: intermediate and Scenario C: IMT-2020 compliant.*

Reference [20] shows a plan for upgrading mobile network to 5G in three Indonesian main cities based on model proposed in [10], forecasting number of users using Bass modeling, and using four base station classes for deployment: 4G LTE-A Macro Base Stations, 5G mmW Metro Base Stations, 5G mmW Pico Base Stations and WiFi IEEE 802.11ac Access Points. It provides discounted CAPEX and OPEX costs.

Reference [21] proposes a framework for indoor small cells deployment in campus, leveraging network slicing for different use cases (eMBB, eMTC, ULLC) operated by micro operators (uOs) that act as a neutral host for Mobile Network Operators (MNOs). The analysis has into account technical benefits: reliability, flexibility, scalability, security, indoor navigation, mobility management, power usage, expansion to new spectrum and network slicing capabilities as well as costs: TCO, CAPEX, OPEX. It considers different type of campus end users requirements for customized services and adequate network slicing.

Reference [22] presents a techno-economic study for optimal deployment of 5G front hauling in large urban areas comparing CPRI (Common Public Radio Interface) and DSP (Digital Signal Processing) assisted radio over fiber channel aggregation, using an algorithm called OTS (Optimal Topology Search) to find optimal locations for Remote Radio Heads (RRH) and diminish deployment costs function.

In [23] a techno-economic comparison of front-/backhauling solutions is presented for a Macro LTE-A Layer with a Pico Layer of small cells deployment, considering wireless point-to-point and wireless point-to-multipoint links to connect eNodeBs to Aggregation Routers, Small Cells or Remote Radio Units (RRUs) to eNodeBs, as well as another alternative with fiber optic links to connect eNodeBs to Aggregation Routers, and RRUs to eNodeBs. The analysis provides TCO comparison between mentioned front-/backhauling options, including CAPEX, IMPEX (Implementation Expenses) and discounted OPEX.

Reference [24] shows a techno-economic analysis of a 5G PPP SESAME C-RAN architecture with edge computing nodes and small cells for immersive video services (IVS) in Crowed Events (CE), following the Neutral Host business model approach by H2020 5GCity project. It uses a techno-economic methodology based on previous TONIC and ECOSYS

European funded projects [1, 2], providing economic output parameters: NPV, IRR, TCO, CAPEX, OPEX and calculating Payback Period.

### B. Conclusions of the literature review

The literature review of relevant papers in previous section demonstrates that there is interest in techno-economic modeling for 5G solutions.

Most of the selected techno-economic papers have into account mainly cost modeling with economic cost output parameters such as CAPEX, OPEX, IMPEX and TCO. Only a few models consider Benefit or business viability output parameters such as NPV, ROI or IRR [5, 8, 7, 10, 24].

Nearly all papers consider enhanced Mobile Broad Band (eMBB) use case. Only 4 out of 18 models take into account massive Machine-Type Communications (mMTC) and 7 out of 18 study Ultra Low Latency Communications (ULLC).

There are multiple technical parameters such as Analysis, Inspection and Simulation KPIs that should be evaluated as defined by 5G PPP, which are not taken into account by most of literature models, along with the fact that there are more technical parameters to be considered and only one model of the literature includes a score for overall technical performance [19].

Only three models show possibility of automatable assessment [7, 15, 22].

All reviewed models use a mobile network operator deployment approach, but only one considers additionally end users` requirements for customized services [21].

No model includes an end customer´s approach to help, for instance, enterprise customers` decision-making to select optimal 5G architectural solutions among different proposals from one or several mobile network operators.

In conclusion, all models in the literature are elaborated specifically for a reduced number of scenarios. They lack generalization capabilities that allow their adaptation to different use cases and evolving 5G architectures. Reviewed models are not flexible enough to integrate new technical and economic parameters as well as other stakeholders´ perspectives, and do not allow agile assessment through automation.

### C. Research problem

In the light of the above, there is a wide nature of technical parameters to be assessed and different 5G architectural options to be taken into account in order to evaluate technical feasibility of 5G solutions. There is also a set of economic output parameters that must consider costs and benefits for an effective evaluation of economic feasibility. On the other hand, all stakeholders' perspectives should be contemplated, and automation should be enabled for agile assessment.

Hence, it makes sense to delve into the research for new techno-economic models that cover all these needs in a more flexible and generalizable way than current models in the literature. These new models should be able to integrate the multiple and evolving technical parameters, the increasing variety of architectural options for 5G solutions, all stakeholders´ perspectives and automation capabilities.

In order to progress in that research using a more structured method to identify current gaps in the literature, the characteristics of a theoretical 5G techno-economic assessment reference model are proposed in the following section.

## III. Characteristics of the theoretical 5G Techno-Economic Assessment Reference Model

Based on the review of the literature, the characteristics of a theoretical 5G techno-economic assessment reference model, which considers all stakeholders' perspectives, as well as technical and economic feasibility of any 5G current and future architecture as well as automation capabilities for agile assessment, are proposed in this section as 15 characteristics named C1-C15. Acronyms are included for easy identification in Table I (RT, NT, MB, MT, LL, IN, AN, SI, EC, DC, CR, CB, OT, MP, AU):

- C1. Assessment of both RAN and Transport Networks (Acronym RT).
- C2. Business viability NPV and TCO (Acronym NT).
- C3. Evaluation of eMBB (enhanced Mobile Broad Band) (Acronym MB).
- C4. Evaluation of massive machine-type communications mMTC (Acronym MT).
- C5. Evaluation of ultra reliable low latency communications (URLLC) (Acronym LL).
- C6. Evaluation of Inspection KPIs of 5G PPP (Acronym IN).
- C7. Evaluation of Analytical KPIs of 5G PPP (Acronym AN).
- C8. Evaluation of Simulation KPIs of 5G PPP (Acronym SI).
- C9. Economic assessment providing OPEX, CAPEX, Revenues, ARPU as output parameters (Acronym EC).
- C10. Demand and Capacity assessment (Acronym DC).
- C11. Evaluation of different degrees of centralization RAN or CRAN (Acronym CR).
- C12. Consideration of Cost-Benefit analysis for decision (Acronym CB).
- C13. Overall Technical Performance (Acronym OT).
- C14. Multi-Perspective: including all stakeholders´ perspective (not only Mobile Network Operator deployment perspective) (Acronym MP).
- C15. Capabilities to Automate Assessment (Acronym AU).

The following section introduces the classification of the reviewed models in the literature considering their compliance with these 15 characteristics of the theoretical reference model.

## IV. Classification of models in the literature

This section shows the classification of the literature models reviewed in previous section II, according to their fulfilment of the characteristics of the theoretical 5G techno-economic reference model, presented in section III.

Table I shows the characteristics (C1-C15) in columns with their acronyms for the Theoretical Model (row 1) and the 18 relevant models in the literature (in following rows). Table I assigns '1' value to a cell if the model in that row is compliant with the characteristic of the corresponding column, and no value ('0' value) if the model does not fulfil that characteristic. The summation of values for each row appears in the penultimate column (Total). The last column (Compliance) provides the overall degree of compliance of the reviewed model in that row.

The ranking in Table I shows that the most compliant model is Roblot et al., 2019 [12], that obtains 8 points out of 15, which means a degree of overall compliance of 53,33% with the theoretical 5G techno-economic assessment reference model. Therefore, there is a gap for improvement at least of 46,66% to reach full compliance with future models.

The characteristics to be improved considering TOP 6 reviewed models are C2 (NT: Business viability NPV and TCO), C6 (IN: Evaluation of Inspection KPIs of 5G PPP), C7 (AN: Evaluation of Analytical KPIs of 5G PPP), C12 (CB: Cost-Benefit analysis for decision), C13 (OT: Overall Technical Performance), C14 (MP: Multi-Perspective: including all stakeholders' perspective -not only Mobile Network Operator deployment perspective-) and C15 (AU: Capabilities to Automate Assessment).

## V. Conclusions and Future work

This article has proposed the characteristics a theoretical 5G Techno-Economic Assessment reference model should have (Section III). After reviewing relevant papers of the literature (Section II), a classification of them has been shown based on their degree of compliance with the characteristics of the theoretical reference model (Section IV). The most compliant model of the literature presents 53,33% of compliance, thus leaving a gap of 46,66% for further improvement in future models. From the analysis of the gap in characteristics for the TOP 6 reviewed models in the ranking, future improvement should be focused on providing business viability outputs (NPV and TCO), evaluation of 5G PPP inspection and analytical KPIs, providing full spectrum of economic outputs (CAPEX, OPEX, Revenues, ARPU, NPV, IRR, TCO, Payback Period, …) to consider Cost-Benefit analysis for decision, evaluation of overall technical performance, integration of multiple stakeholders' perspectives (not only mobile network operators deployment perspective) and automatic assessment capabilities in order to achieve effective and agile 5G techno-economic assessment.

Author´s future work in this direction will consider using a universal techno-economic model called UTEM, created and developed by the author, for effective and agile 5G techno-economic assessment, including not only network deployment perspective but also customers` requirements and other stakeholders´ in order to select the most adequate technical solution considering technical and economic feasibility [1, 2].

UTEM model is a universal, flexible, scalable and generalizable model that allows to industrialize techno-economic assessment for agile decision-making in multiple domains, considering all market players perspectives, also suitable for technological consulting and currently available for all industry stakeholders under specific license of use [1].

**Acknowledgments**

The author thanks all academy and industry contributors for the nascent 5G techno-economic literature, as they have made possible this work to review current literature and identify improvement gaps for future 5G techno-economic models, which are key for our societies to get full benefit of 5G technology and ecosystem.

*Table I. Classification of models in the literature according to the characteristics established for the theoretical model.*

| Models | Characteristics of the Theoretical 5G Techno-economic Assessment Reference Model ||||||||||||||| Total | Compliance |
|---|---|---|---|---|---|---|---|---|---|---|---|---|---|---|---|---|---|
| | C1 RT | C2 NT | C3 MB | C4 MT | C5 LL | C6 IN | C7 AN | C8 SI | C9 EC | C10 DC | C11 CR | C12 CB | C13 OT | C14 MP | C15 AU | | |
| Theoretical Model | 1 | 1 | 1 | 1 | 1 | 1 | 1 | 1 | 1 | 1 | 1 | 1 | 1 | 1 | 1 | 15 | 100% |
| Roblot et al., 2019 [12] | 1 | | 1 | 1 | 1 | | | 1 | 1 | 1 | 1 | | | | | 8 | 53,33% |
| Maternia et al., 2018 [4] | | | 1 | 1 | 1 | 1 | 1 | 1 | 1 | | | | | | | 7 | 46,66% |
| mmMAGIC, 2017 [8] | 1 | 1 | 1 | | 1 | | | | | 1 | 1 | 1 | | | | 7 | 46,66% |
| J.R. Martín et al., 2019 [19] | | | 1 | 1 | 1 | | | 1 | 1 | 1 | | | 1 | | | 7 | 46,66% |
| Mesogiti et al., 2020 [15] | 1 | | 1 | | 1 | | | | 1 | 1 | 1 | | | 1 | | 7 | 46,66% |
| Oughton et al., 2019 [7] | 1 | 1 | 1 | | | | | 1 | 1 | 1 | | | | 1 | | 7 | 46,66% |
| Smail et al., 2017 [10] | 1 | 1 | 1 | | | | | | 1 | 1 | | 1 | | | | 6 | 40,00% |
| Bouras et al., 2019 [16] | 1 | | 1 | | | | | | 1 | 1 | 1 | 1 | | | | 6 | 40,00% |
| Neokosmidis et al., 2019 [24] | | 1 | 1 | | 1 | | | | 1 | 1 | | 1 | | | | 6 | 40,00% |
| Walia et al., 2017 [21] | | | 1 | 1 | 1 | | | | 1 | 1 | | | | 1 | | 6 | 40,00% |
| Yaghoubi et al., 2018 [5] | 1 | 1 | 1 | | | | | | 1 | 1 | | 1 | | | | 6 | 40,00% |
| Bouras et al., 2015 [13] | 1 | | 1 | | | | | 1 | 1 | 1 | | | | | | 5 | 33,33% |
| Bouras et al., 2016 [14] | 1 | | 1 | | | | | | 1 | 1 | 1 | | | | | 5 | 33,33% |
| Kolydakis and Tomkos, 2014 [23] | 1 | | 1 | | | | | | 1 | 1 | | | | | | 4 | 26,66% |
| Nikolikj and Janevski, 2014 [11] | | | 1 | | | | | | 1 | 1 | | 1 | | | | 4 | 26,66% |
| Kusuma and Suryanegara, 2019 [20] | 1 | | 1 | | | | | | 1 | 1 | | | | | | 4 | 26,66% |
| Bouras et al., 2018 [17] | | | 1 | | | | | | 1 | 1 | | | | | | 3 | 20,00% |
| Arévalo et al., 2018 [22] | | | 1 | | | | | | 1 | | | | | 1 | | 3 | 20,00% |